\documentclass[twocolumn,twoside,slac_two]{revtex4}
\usepackage{graphicx}
\usepackage{fancyhdr}
\def\MET{{\mbox{$E\kern-0.57em\raise0.19ex\hbox{/}_{T}$}}}
\def\met{{\mbox{$E\kern-0.57em\raise0.19ex\hbox{/}_{T}$}}}
\def\DZ{D\O\ }
\def\DZero{D\O\ }
\def\Dzero{D\O\ }

\def\ifb{~fb$^{-1}$}
\def\pp{$p\bar{p}$}

\def\WH{$WH\rightarrow \ell\nu b\bar{b}$}
\def\WHt{$WH\rightarrow \tau\nu b\bar{b}$}
\def\lmet{$WH\rightarrow \ell\kern-0.45em\raise0.19ex\hbox{/} \nu b\bar{b}$}
\def\ZH{$ZH\rightarrow \nu\bar{\nu} b\bar{b}$}
\def\ZHll{$ZH\rightarrow \ell^+ \ell^- b\bar{b}$}

\def\www{$WH \rightarrow WW^{+} W^{-}$}

\def\hww{$H\rightarrow W^+ W^-$}

\def\tevE{$\sqrt{s}=1.96$~TeV}

\begin{document}

\title{Combined SM Higgs Limits at the Tevatron}

%

\author{N.~Krumnack on behalf of the CDF and D\O\ collaborations}
\affiliation{Iowa State University, Department of Physics and Astronomy, Physics Hall, Ames, IA 50011-3160, USA}

\begin{abstract}
We combine results from CDF and D\O\ on direct searches for a standard
model (SM) Higgs boson ($H$) in \pp~collisions at the Fermilab
Tevatron at $\sqrt{s}=1.96$~TeV. Compared to the previous Higgs
Tevatron combination, more data and new channels ($WH \rightarrow \tau
\nu b \bar{b}$, $VH \rightarrow \tau \tau b \bar{b}/jj \tau\tau$, $VH
\rightarrow jj b \bar{b}$, $t \bar{t} H \rightarrow t \bar{t} b
\bar{b}$) have been added.  Most previously used channels have been
reanalyzed to gain sensitivity. We use the latest parton distribution
functions and $gg \rightarrow H$ theoretical cross sections when
comparing our limits to the SM predictions. 
 With 2.0-3.6\ifb\ of data
analyzed at CDF, and 0.9-4.2\ifb\ at D\O, the 95\% C.L. upper limits
on Higgs boson production are a factor of 2.5~(0.86) times the SM
cross section for a Higgs boson mass of $m_{H}=$115~(165)~GeV/c$^2$.
Based on simulation, the corresponding median expected upper limits
are 2.4 (1.1). 
The mass range excluded at 95\% C.L. for a SM
Higgs has been extended to  $160<m_{H}<170$~GeV/c$^{2}$.
\end{abstract}

\maketitle

\thispagestyle{fancy}


\section{Introduction}

The search for a mechanism for electroweak symmetry breaking, and in
particular for a standard model (SM) Higgs boson has been a major goal
of particle physics for many years, and is a central part of the
Fermilab Tevatron physics program. Both the CDF and \Dzero experiments
are reporting new combinations~\cite{CDFhiggs,DZhiggs} of multiple
direct searches for the SM Higgs boson.  The new searches include more
data and improved analysis techniques compared to previous analyses.
The sensitivities of these new combinations significantly exceed
previous work~\cite{CDFHiggsICHEP,DZHiggsICHEP}.  The most recent
Tevatron Higgs combination~\cite{TevHiggsICHEP} only included channels
seeking Higgs bosons of masses between 155 and 200~GeV/$c^2$, and the
most recent combination over the entire mass range 100-200~GeV/$c^2$
was reported in April 2008~\cite{tev-apr-2008}.

In this note, we combine the most recent results of all such searches
in \pp~collisions at~\tevE.  The analyses combined here seek signals
of Higgs bosons produced in associated with vector bosons
($q\bar{q}\rightarrow W/ZH$), through gluon-gluon fusion
($gg\rightarrow H$), and through vector boson fusion (VBF) ($q\bar{q}\rightarrow q^{\prime}\bar{q}^{\prime}H$)
corresponding to integrated luminosities ranging from 2.0-3.6\ifb~at
CDF and 0.9-4.2\ifb~at D\O. The Higgs boson decay modes studied are
$H\rightarrow b{\bar{b}}$, $H\rightarrow W^+W^-$, $H\rightarrow
\tau^+\tau^-$ and $H\rightarrow \gamma\gamma$.

To simplify the combination, the searches are separated into 75
mutually exclusive final states (23 for CDF and 52 for D\O; see
Table~\ref{tab:cdfacc} and ~\ref{tab:dzacc}) referred to as
``analyses'' in this note.  The selection procedures for each analysis
are detailed in Refs.~\cite{cdfWH} through~\cite{dzttH}, and are
briefly described below.

\begin{table}[htbp]
\begin{center}
\caption{\label{tab:cdfacc}Luminosity, explored mass range and references
for the different processes
and final state ($\ell=e, \mu$) for the CDF analyses}
\begin{tabular}{|l|c|c|c|}\hline
Channel & Lumi. & $m_H$ range & Ref. \\ 
& (fb$^{-1}$) & (GeV/$c^2$) & \\ \hline
$WH\rightarrow \ell\nu b\bar{b}$        & 2.7  & 100-150 & \cite{cdfWH} \\
$ZH\rightarrow \nu\bar{\nu} b\bar{b}$       & 2.1  & 105-150 & \cite{cdfZH} \\
$ZH\rightarrow \ell^+\ell^- b\bar{b}$   & 2.7  & 100-150 & \cite{cdfZHll} \\
$H\rightarrow W^+ W^-$  & 3.6  & 110-200 & \cite{cdfHWW} \\
$WH \rightarrow WW^+ W^- \rightarrow \ell^\pm\nu \ell^\pm\nu$ & 3.6  & 110-200 & \cite{cdfHWW} \\
$H$ + $X\rightarrow \tau^+ \tau^- $ + 2 jets                  & 2.0  & 110-150 & \cite{cdfHtt} \\
$WH+ZH\rightarrow jjb{\bar{b}}$                               & 2.0  & 100-150 & \cite{cdfjjbb} \\ \hline
\end{tabular}
\end{center}
\end{table}

\begin{table}[htbp]
\caption{\label{tab:dzacc}Luminosity, explored mass range and references 
for the different processes
and final state ($\ell=e, \mu$) for the D\O\ analyses
}
\begin{center}
\begin{tabular}{|l|c|c|c|}\hline
Channel & Lumi. & $m_H$ range & Ref. \\ 
& (fb$^{-1}$) & (GeV/$c^2$) & \\ \hline
$WH\rightarrow \ell\nu b\bar{b}$             & 2.7  & 100-150 & \cite{dzWHl} \\
$WH\rightarrow \tau\nu b\bar{b}$             & 0.9  & 105-145 & \cite{dzWHt} \\
$VH\rightarrow \tau\tau b\bar{b}/q\bar{q} \tau\tau$           & 1.0  & 105-145 & \cite{dzWHt} \\
$ZH\rightarrow \nu\bar{\nu} b\bar{b}$                   & 2.1  & 105-145 & \cite{dzZHv} \\
$ZH\rightarrow \ell^+\ell^- b\bar{b}$        & 2.3  & 105-145 & \cite{dzZHll} \\
$WH \rightarrow WW^+ W^- \rightarrow \ell^\pm\nu \ell^\pm\nu$ & 1.1  & 120-200 & \cite{dzWWW} \\
$H\rightarrow W^+ W^- \rightarrow \ell^\pm\nu \ell^\mp\nu$    & 3.0--4.2  & 115-200 & \cite{dzHWW}\\
$H \rightarrow \gamma \gamma$                                 & 4.2  & 100-150 & \cite{dzHgg} \\ 
$t \bar{t} H \rightarrow t \bar{t} b \bar{b}$ & 2.1  & 105-145 & \cite{dzttH} \\ \hline
\end{tabular}
\end{center}
\end{table}

\section{Acceptance, Backgrounds and Luminosity}  

Event selections are similar for the corresponding CDF and D\O\ analyses.
For the case of \WH, an isolated lepton ($\ell=$ electron or muon) and 
two jets are required, with one or more $b$-tagged jet, i.e., identified 
as containing a weakly-decaying $B$ hadron.  
Selected events must also display a significant imbalance 
in transverse momentum
(referred to as missing transverse energy or \met).  Events with more than one
isolated lepton are vetoed.  
For the D\O\ \WH\ analyses, two and three jet events are analyzed separately, and 
in each of these samples
two non-overlapping $b$-tagged samples are
defined, one being a single ``tight'' $b$-tag (ST) sample, and the other a 
double ``loose'' $b$-tag (DT) sample. The tight and loose $b$-tagging criteria
are defined with respect to the mis-identification 
rate that the $b$-tagging algorithm yields for light quark or gluon jets 
(``mistag rate'') typically $\le 
0.5\%$ or $\le 1.5\%$, respectively.  
The final variable is a neural network output which takes as input seven kinematics
variables and a matrix element discriminant for the 2 jet sample, while for the
3 jet sample the dijet invariant mass is used.
In this combination, we add a new analysis
\WHt\ in which the $\tau$ is identified through its hadronic decays.
this analysis is sensitive to $ZH \rightarrow \tau \met b \bar{b}$
 as well in those cases where a $\tau$ fails to be identified.
The analysis is carried out according to the type of reconstructed 
$\tau$ and  is also 
separated into two and three jets with DT events only.
It uses the dijet invariant
mass of the $b\bar{b}$ system as discriminant variable.

For the CDF \WH\ analyses, the events are grouped into six categories.  In addition
to the selections requiring an identified lepton, events with an isolated track
failing lepton selection requirements are grouped into their own categories.  
This provides some acceptance for single prong tau decays.
Within the
lepton categories there 
are three $b$-tagging categories -- two tight $b$-tags (TDT), one tight $b$-tag
and one loose $b$-tag (LDT), and a single, tight, $b$-tag (ST).  In each category, two discriminants
are calculated for each event.  One neural network discriminant is trained at each
$m_H$ in the test range, separately for each category.  A second discriminant is a 
boosted decision tree, featuring not only event kinematic and $b$-tagging observables,
but matrix element discriminants as well.  These two discriminants are then combined
together using an evolutionary neural network~\cite{NEAT} to form a single discriminant with optimal
performance.

For the \ZH\ analyses, the selection is
similar to the $WH$ selection, except all events with isolated leptons are vetoed and
stronger multijet background suppression techniques are applied. 
Both CDF and D\O\  analyses use a track-based missing transverse momentum calculation
as a discriminant against false \met .  
There is a sizable fraction of \WH\ signal in which the lepton is undetected,
that is selected in the \ZH\ samples,  so these analyses are also refered to as
$VH \rightarrow \met b \bar{b}$.
The CDF  analysis uses three non-overlapping samples of 
events (TDT, LDT and ST as for $WH$) while D\O\
uses a sample of events having one tight $b$-tag jet and one loose $b$-tag jet.
CDF used
neural-network discriminants as the final variables, while D\O\ 
uses boosted decision trees as advanced analysis technique.

The \ZHll\ analyses require two isolated leptons and
at least two jets. They   use non-overlapping samples of
events with one tight $b$-tag and two loose $b$-tags. 
For the D\O\ analysis  neural-network and boosted decision trees
discriminants are the 
final variables for setting  limits (depending on the sub-channel), 
while  CDF uses the output
of a 2-dimensional neural-network.  
CDF corrects jet energies for \met\ using a neural network approach.
In this analysis  also the events are
divided  into three tagging
categories: tight double tags, loose double tags, and single tags.

For the \hww~analyses, signal events are characterized by
a large \met~and two opposite-signed, isolated leptons.  The presence of
neutrinos in the final state prevents the reconstruction of the
candidate Higgs boson mass. 
 D\O\ selects events containing electrons and muons,
dividing the data sample into three final states:
$e^+e^-$, $e^\pm \mu^\mp$, and $\mu^+\mu^-$.
CDF separates the \hww\ events in five non-overlapping samples,
labeled ``high $s/b$'' and ``low $s/b$'' for the lepton selection categories, and
also split by the number of jets: 0, 1, or 2+ jets.  The sample with two or more jets
is not split into low $s/b$ and high $s/b$ lepton categories.   The division of events
into jet categories allows the analysis discriminants to separate three different categories of
signals from the backgrounds more effectively.  The signal production mechanisms considered are
$gg\rightarrow H\rightarrow W^+W^-$, $WH+ZH\rightarrow jjW^+W^-$, and the vector-boson
fusion process.  The final discriminants are neural-network outputs for D\O\, and
neural-network output
including likelihoods constructed from matrix-element probabilities (ME) as input
to the neural network, for CDF, in the 0-jet bin, else the ME are not used.
 All analyses in this channel have
been updated with more data and analysis improvements.

The CDF collaboration also contributes an analysis searching for Higgs bosons decaying
to a tau lepton pair, in three separate production channels:
direct $p \bar{p} \rightarrow H$ production, associated $WH$ or $ZH$ production,
or vector boson production with $H$ and forward jets in the final state.
In  this analysis, the final variable for setting  limits is
a combination of several neural-network discriminants.

D\O\ also
contributes a new analysis 
for the final state $\tau
\tau$ jet jet, which is sensitive to the $VH\rightarrow jj \tau \tau$, $ZH 
\rightarrow \tau \tau b \bar{b}$, VBF and gluon gluon fusion
(with two additional jets) mechanisms. It
uses a neural network output  as discriminant variable. 

The CDF collaboration introduces a new all-hadronic channel, $WH+ZH\rightarrow jjb{\bar{b}}$ for this
combination.  Events are selected with four jets, at least two of which are $b$-tagged
with the tight $b$-tagger.  The large QCD backgrounds are estimated with the use of
data control samples, and the final variable is a matrix element signal probability
discriminant.

The D\O\ collaboration
contributes three \www\ analyses, where the associated $W$ boson and
the $W$ boson from the Higgs boson decay which has the same charge are required 
to decay leptonically,
thereby defining three like-sign dilepton final states 
($e^\pm e^\pm$, $e^\pm \mu^\pm$, and $\mu^{\pm}\mu^{\pm}$)
containing all decays of the third
$W$ boson. In  this analysis, which has not been updated for this combination,
the final variable is a likelihood discriminant formed from several
topological variables.   CDF contributes a \www\ analysis using a selection of like-sign
dileptons and a neural network to further purify the signal.
D\O\ also contributes an analysis searching for direct Higgs boson production
decaying to a photon pair in 4.2 fb$^{-1}$ of data.
In  this analysis, the final variable is the invariant mass of the two-photon system.

Another new search from D\O\ is
included in this combination, namely the search for
$t \bar{t} H \rightarrow t \bar{t} b \bar{b}$. Here the samples are analyzed 
independently according to the
number of $b$-tagged jets (1,2,3, i.e. ST,DT,TT) 
and the total number of jets (4 or 5).
The total transverse energy of the reconstructed objects ($H_T$) is used
as discriminant variable.

All Higgs boson signals are simulated using \textsc{PYTHIA}~\cite{pythia}, and 
\textsc{CTEQ5L} or \textsc{CTEQ6L}~\cite{cteq} leading-order (LO)
parton distribution functions.  
The $gg\rightarrow H$ production cross section is calculated at NNLL in
QCD and also includes two-loop electroweak
effects; see Refs.~\cite{anastasiou,grazzinideflorian} and references therein for 
the different steps of these calculations.   The newer calculation
includes a more thorough treatment of higher-order radiative corrections, particularly those involving
$b$ quark loops.  The $gg\rightarrow H$ production cross section depends strongly on the PDF set chosen and the
accompanying value of $\alpha_s$.  The cross sections used 
here are calculated with the MSTW 2008 NNLO PDF set~\cite{mstw2008}.   The new $gg\rightarrow H$ cross sections
supersede those used in the update of Summer 2008~\cite{TevHiggsICHEP,nnlo1,aglietti}, which had a simpler
treatment of radiative corrections and used the older MRST 2002 PDF set~\cite{mrst2002}.  The Higgs boson
production cross sections used here
are listed in~\cite{TevHiggs} (originally from~\cite{grazzinideflorian}).  
We include all significant Higgs production modes in the high mass search: besides gluon-gluon 
fusion through a virtual top quark loop  (ggH), 
we include production in association 
with a $W$ or $Z$ vector boson  (VH)~\cite{nnlo2,Brein,Ciccolini}, 
and vector boson  fusion (VBF)~\cite{nnlo2,Berger}.

The Higgs boson
decay branching ratio predictions are calculated with HDECAY~\cite{hdecay}.
For both CDF and D\O , events from
multijet (instrumental) backgrounds (``QCD production'') are measured
in data with different methods, in orthogonal samples.
For CDF, backgrounds
from other SM processes were generated using \textsc{PYTHIA},
\textsc{ALPGEN}~\cite{alpgen}, \textsc{MC@NLO}~\cite{MC@NLO}
 and \textsc{HERWIG}~\cite{herwig}
programs. For D\O , these backgrounds were generated using
\textsc{PYTHIA}, \textsc{ALPGEN}, and \textsc{COMPHEP}~\cite{comphep},
with \textsc{PYTHIA} providing parton-showering and hadronization for
all the generators.  These background processes were normalized using either
experimental data or next-to-leading order calculations (from
\textsc{MCFM}~\cite{mcfm} for $W+$ heavy flavor process).

Integrated luminosities, 
and references to the collaborations' public documentation for each analysis
are given in Table~\ref{tab:cdfacc}
for CDF and in Table~\ref{tab:dzacc} for D\O .  
The tables include the ranges of Higgs boson mass ($m_H$) over which
the searches were performed. 

\section{Distributions of Candidates} 

The number of channels combined is quite large, and the number of bins
in each channel is large.  Therefore, the task of assembling
histograms and checking whether the expected and observed limits are
consistent with the input predictions and observed data is difficult.
We therefore provide histograms that aggregate all channels' signal,
background, and data together.  In order to preserve most of the
sensitivity gain that is achieved by the analyses by binning the data
instead of collecting them all together and counting, we aggregate the
data and predictions in narrow bins of signal-to-background ratio,
$s/b$.  Data with similar $s/b$ may be added together with no loss in
sensitivity, assuming similar systematic errors on the predictions.
The aggregate histograms do not show the effects of systematic
uncertainties, but instead compare the data with the central
predictions supplied by each analysis.

The range of $s/b$ is quite large in each analysis, and so
$\log_{10}(s/b)$ is chosen as the plotting variable.  Plots of the
distributions of $\log_{10}(s/b)$ are shown for $m_H=115$ and
165~GeV/c$^2$ in Figure~\ref{fig:lnsb}.  These distributions can be
integrated from the high-$s/b$ side downwards, showing the sums of
signal, background, and data for the most pure portions of the
selection of all channels added together.  These integrals can be seen
in Figure~\ref{fig:integ}.  

 \begin{figure}[htbp]
 \begin{centering}
 \includegraphics[width=0.25\textwidth]{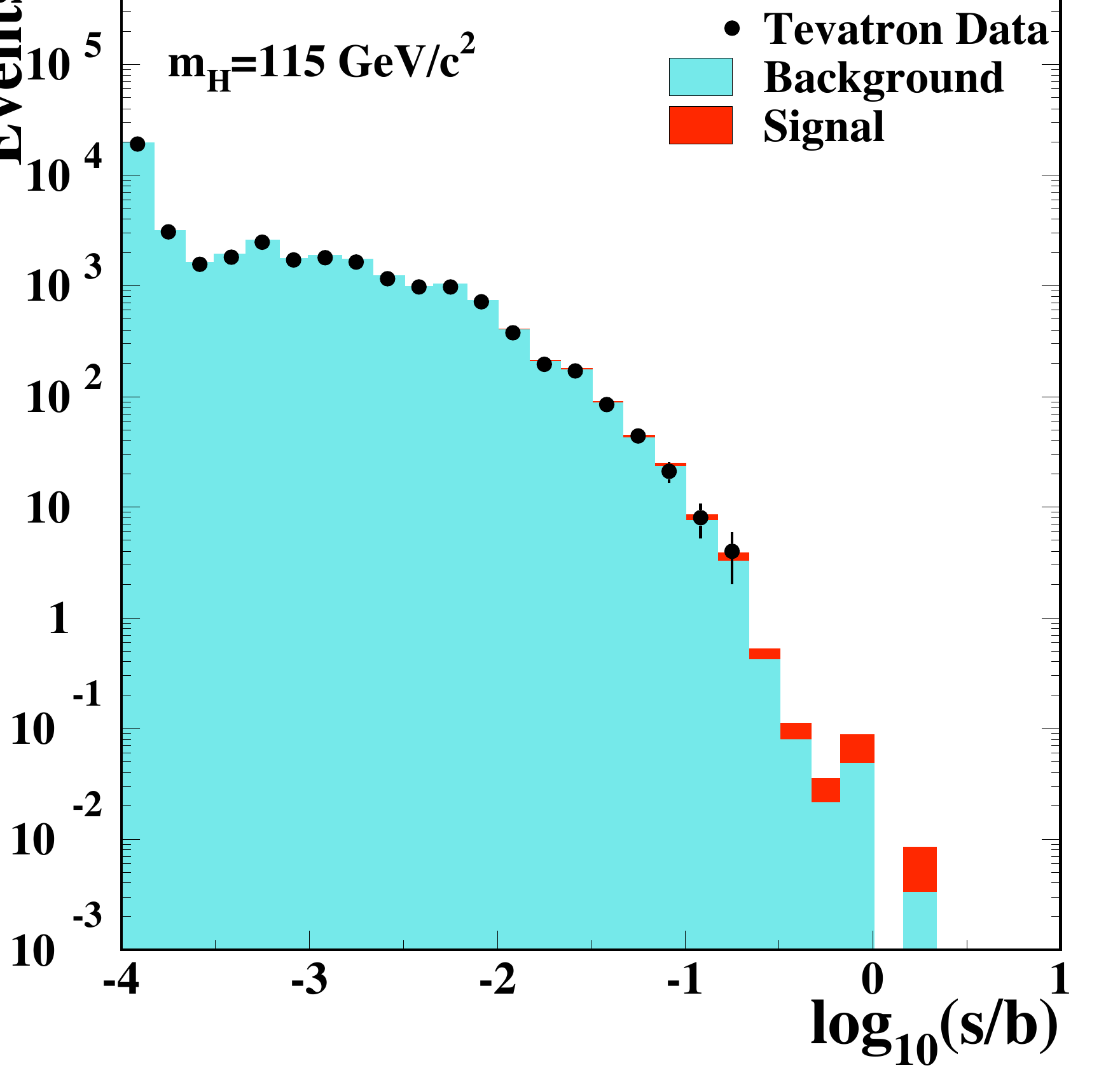}\includegraphics[width=0.25\textwidth]{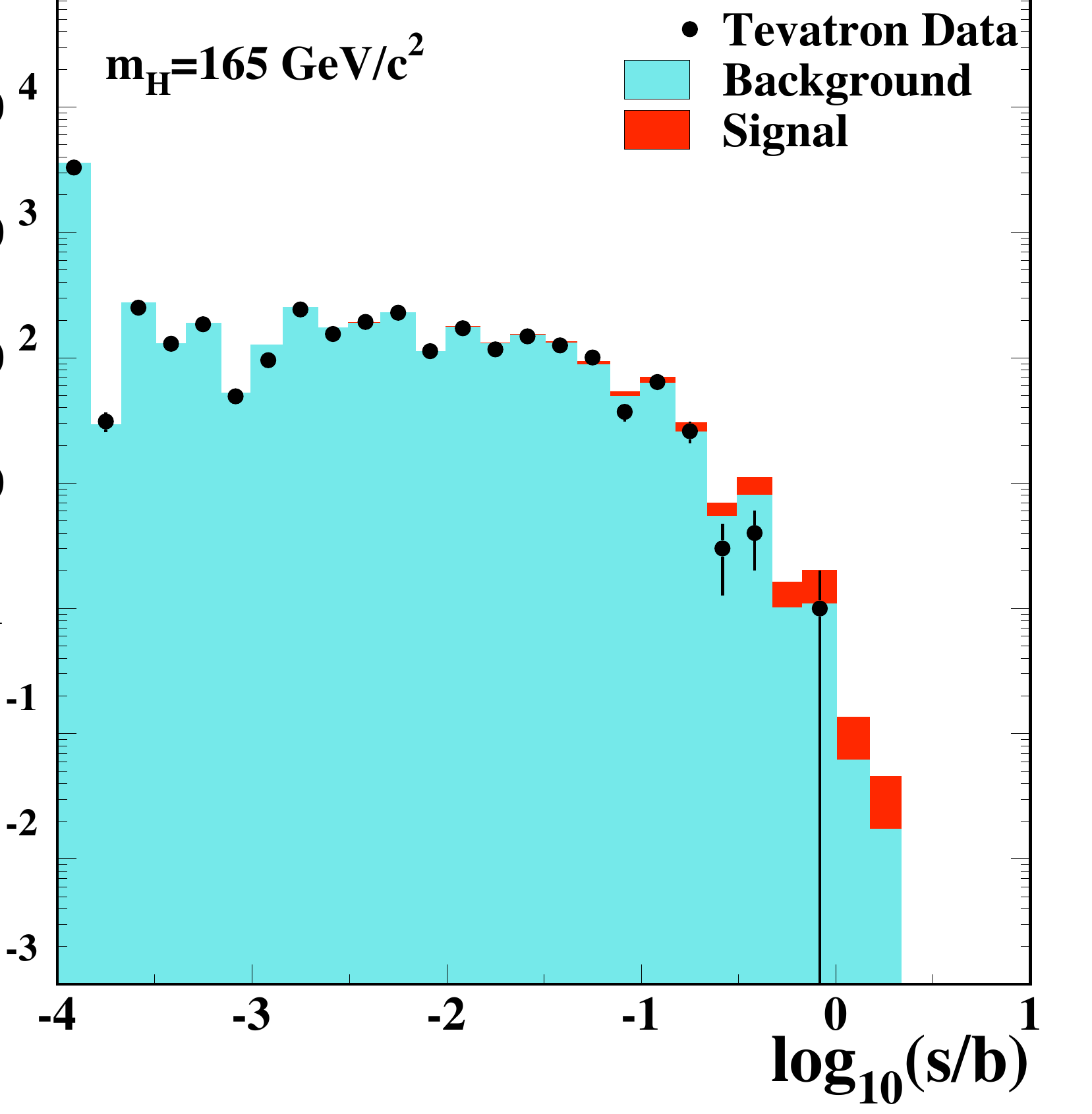}
 \caption{
 \label{fig:lnsb} Distributions of $\log_{10}(s/b)$, for the data from all contributing channels from
CDF and D\O, for Higgs boson masses of 115 and 165~GeV/$c^2$.  The
data are shown with points, and the signal is shown stacked on top of
the backgrounds.  Underflows and overflows are collected into the
bottom and top bins. }
 \end{centering}
 \end{figure}

 \begin{figure}[htbp]
 \begin{centering}
 \includegraphics[width=0.25\textwidth]{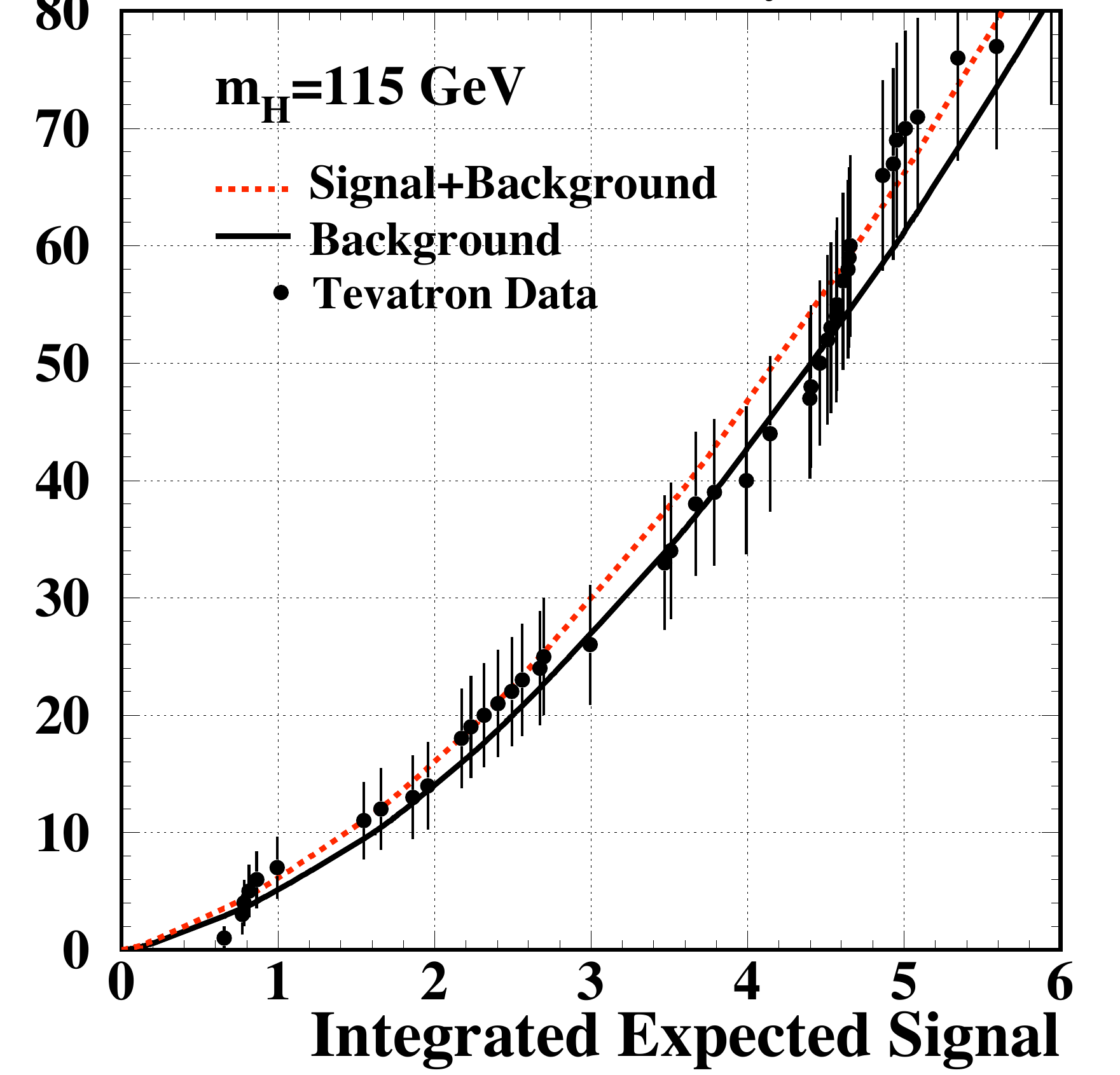}\includegraphics[width=0.25\textwidth]{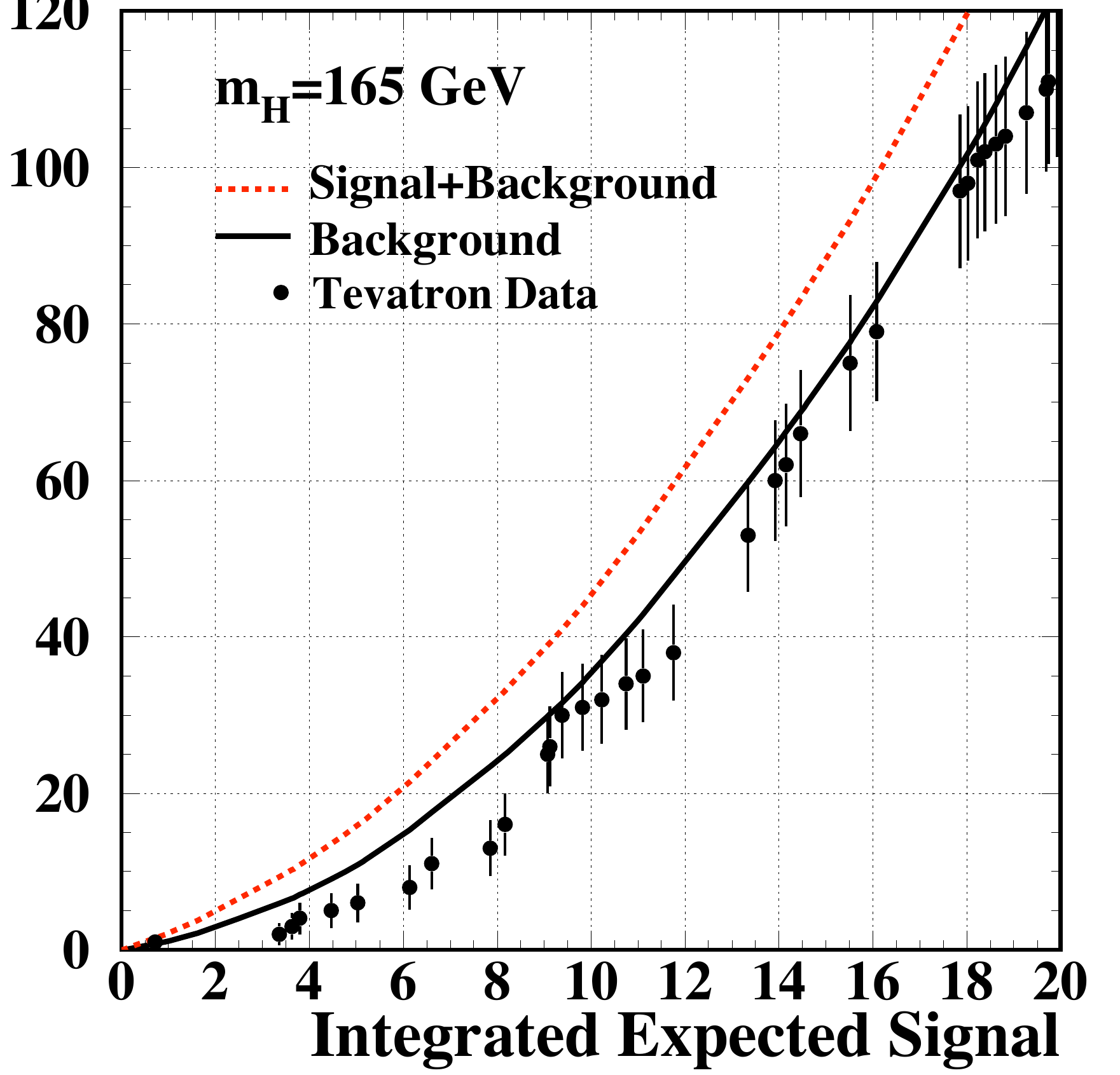}
 \caption{
 \label{fig:integ} Integrated distributions of $s/b$, starting at the high $s/b$ side.  The total signal+background
and background-only integrals are shown separately, along with the
data sums.  Data are only shown for bins that have 
data events in them.}
 \end{centering}
 \end{figure}

\section{Combining Channels} 

To gain confidence that the final result does not depend on the
details of the statistical formulation, 
we perform two types of combinations, using the
Bayesian and  Modified Frequentist approaches, which give similar results
(within 10\%).
Both methods rely on distributions in the final discriminants, and not just on
their single integrated values.  Systematic uncertainties enter as uncertainties on the
expected number of signal and background events, as well
as on the distribution of the discriminants in 
each analysis (``shape uncertainties'').
Both methods use likelihood calculations based on Poisson
probabilities.  Detailed descriptions of the techniques can be found in~\cite{TevHiggs} as well as in~\cite{CDFhiggs} for the Bayesian Approach and in \cite{DZhiggs,pdgstats,pflh} for the Modified Frequentist Approach.

\subsection{Systematic Uncertainties} 

Systematic uncertainties differ
between experiments and analyses, and they affect the rates and shapes of the predicted
signal and background in correlated ways.  The combined results incorporate
the sensitivity of predictions to  values of nuisance parameters,
and include correlations, between rates and shapes, between signals and backgrounds,
and between channels within experiments and between experiments.
More on these issues can be found in the
individual analysis notes~\cite{cdfWH} through~\cite{dzttH} and in the combination note \cite{TevHiggs}.

\section{Combined Results} 

Using the combination procedures outlined in~\cite{TevHiggs}, we extract limits on
SM Higgs boson production $\sigma \times B(H\rightarrow X)$ in
\pp~collisions at $\sqrt{s}=1.96$~TeV for $m_H=100-200$ GeV/c$^2$.
To facilitate comparisons with the standard model and to accommodate analyses with
different degrees of sensitivity, we present our results in terms of
the ratio of obtained limits  to  cross section in the SM, as a function of
Higgs boson mass, for test masses for which
both experiments have performed dedicated searches in different channels.
A value of the combined limit ratio which is less than or equal to one would indicate that
that particular Higgs boson mass is excluded at the
95\% C.L.

The combinations of results of each single experiment, as used in this Tevatron combination,
yield the following ratios of 95\% C.L. observed (expected) limits to the SM 
cross section: 
3.6~(3.2) for CDF and 3.7~(3.9) for D\O\ at $m_{H}=115$~GeV/c$^2$, and 
1.5~(1.6) for CDF and 1.3~(1.8) for D\O\ at $m_{H}=165$~GeV/c$^2$.

The ratios of the 95\% C.L. expected and observed limit to the SM
cross section are shown in Figure~\ref{fig:comboRatio} for the
combined CDF and D\O\ analyses.  Tables with the observed and median expected
ratios  
as obtained by the Bayesian and the $CL_S$ methods are listed in~\cite{TevHiggs}. 
In the following summary we quote only the limits obtained with the 
Bayesian method since they are slightly more conservative (based on the 
expected limits) for the quoted values, but all the equivalent numbers for 
the $CL_S$ method can be retrieved from~\cite{TevHiggs}.
We obtain the observed
(expected) values of 2.5 (2.4) at $m_{H}=115$~GeV/c$^2$, 0.99 (1.1) at
$m_{H}=160$~GeV/c$^2$, 0.86 (1.1) at $m_{H}=165$~GeV/c$^2$, and 0.99
(1.4) at $m_{H}=170$~GeV/c$^2$.  We exclude at the 95\% C.L. the
production of a standard model Higgs boson with mass between 160 and
170 GeV/c$^2$.  This result is obtained with both Bayesian and $CL_S$
calculations.

\begin{figure}[htbp]
\begin{centering}
\includegraphics[width=1.05\linewidth]{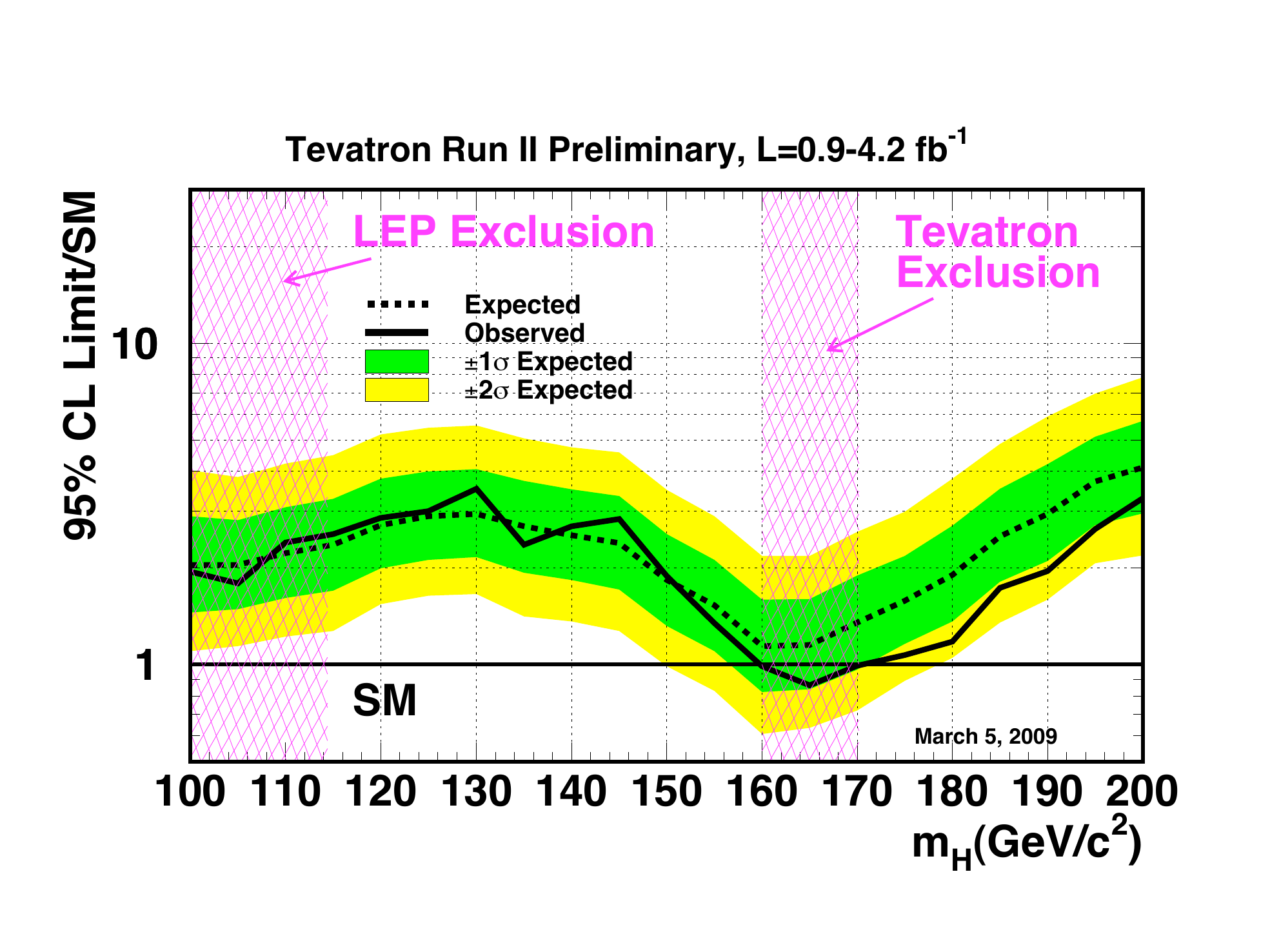}
\caption{
\label{fig:comboRatio}
Observed and expected (median, for the background-only hypothesis)
  95\% C.L. upper limits on the ratios to the
SM cross section, 
as functions of the Higgs boson mass 
for the combined CDF and D\O\ analyses.
The limits are expressed as a multiple of the SM prediction
for test masses (every 5 GeV/$c^2$)
for which both experiments have performed dedicated
searches in different channels.
The points are joined by straight lines 
for better readability.
  The bands indicate the
68\% and 95\% probability regions where the limits can
fluctuate, in the absence of signal. 
The limits displayed in this figure
are obtained with the Bayesian calculation.
}
\end{centering}
\end{figure}

We also show in Figure~\ref{fig:comboLLR-2} the 1-$CL_S$ distribution as a function of 
the Higgs boson mass, at high mass ($\geq 150 $ GeV/$c^2$) which is directly interpreted
as the level of exclusion at 95\% C.L. of our search. Note that this figure is obtained
using the $CL_S$ method. The 90\% C.L. line is also shown on the figure.
We  provide the Log-likelihood ratio (LLR) values for our combined Higgs boson search,
as obtained using the $CL_S$ method in~\cite{TevHiggs}.

 \begin{figure}[htbp]
 \begin{centering}
 \includegraphics[width=\linewidth]{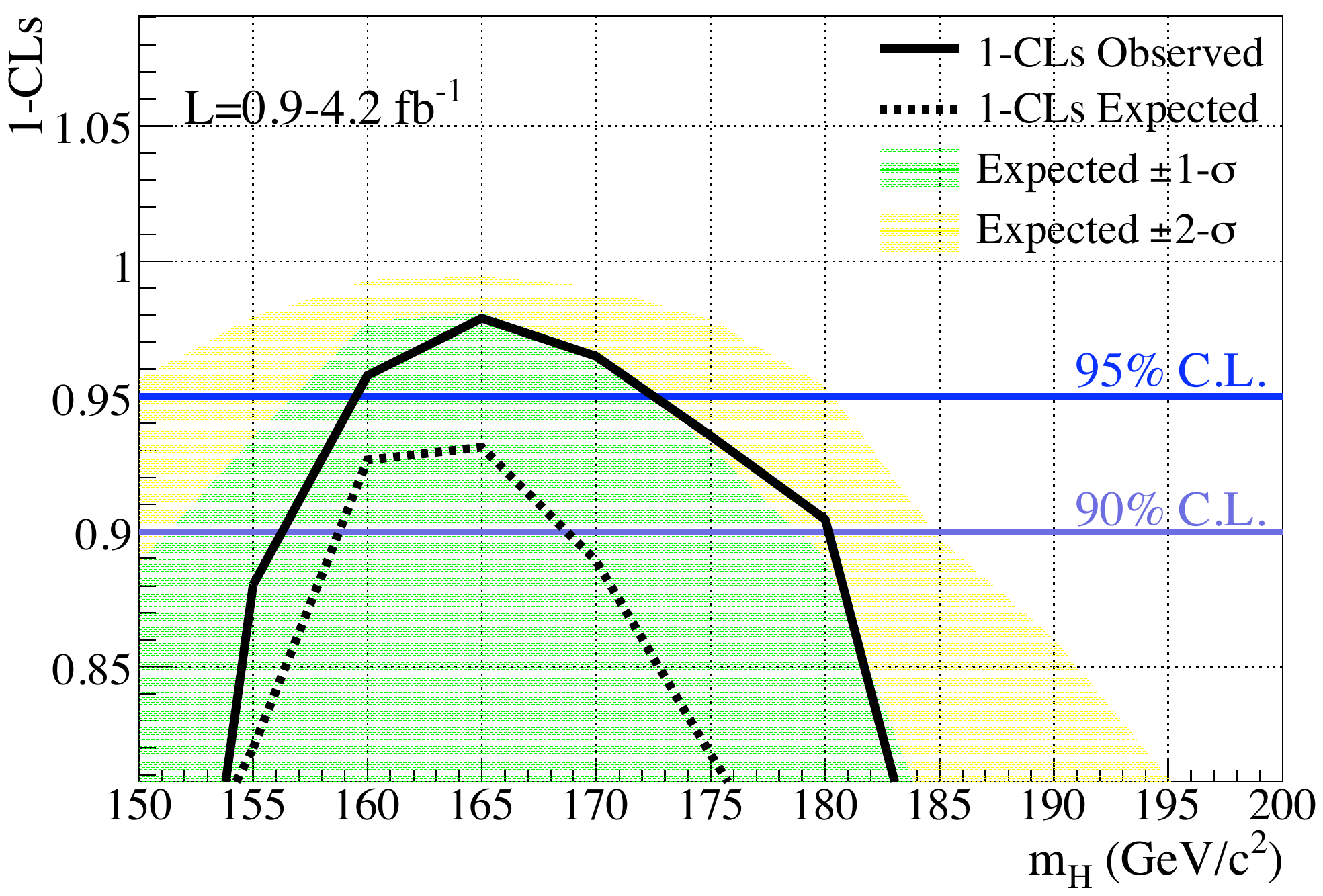}
 \caption{
 \label{fig:comboLLR-2}
 Distributions of 1-$CL_S$ as a function of the Higgs boson mass 
(in steps of 5 GeV/c$^2$), as obtained with $CL_S$ method.
 for the combination of the
 CDF and D\O\ analyses. }
 \end{centering}
 \end{figure}

In summary, we have combined all available CDF and D\O\ results on SM Higgs search,
based on luminosities ranging from 0.9 to 4.2 fb$^{-1}$.
Compared to our previous combination, new channels have been added and
most previously used channels have been
reanalyzed to gain sensitivity. We use the latest parton distribution
functions and $gg \rightarrow H$ theoretical cross sections when
comparing our limits to the SM predictions at high mass.  

The 95\% C.L. upper limits
on Higgs boson production are a factor of 2.5~(0.86) times the SM
cross section for a Higgs boson mass of $m_{H}=$115~(165)~GeV/c$^2$.
Based on simulation, the corresponding median expected upper limits
are 2.4 (1.1). Standard Model branching ratios, calculated as
functions of the Higgs boson mass, are assumed. These results extend
significantly the individual limits of each collaboration and our
previous combination. The mass range excluded at 95\% C.L. for a SM
Higgs has been extended to  $160<m_{H}<170$~GeV/c$^{2}$.
The sensitivity of our combined search is expected to grow substantially in
the near future with the additional luminosity already recorded at the Tevatron
and not yet analyzed, and with additional improvements of our analysis techniques
which will be propagated in the current and future analyses.

\bigskip 

\begin{thebibliography}{9}   

\bibitem{TevHiggs} The TEVNPH Working Group, Combined CDF and \DZ Upper Limits on Standard Model Higgs-Boson Production with up to 4.2 fb$^{-1}$ of Data, FERMILAB-PUB-09-060-E, CDF Note 9713, \DZ Note 5889 (2009).

\bibitem{CDFhiggs} CDF Collaboration, ``Combined Upper Limit on
Standard Model Higgs Boson Production for Winter 2009'', CDF Conference Note 9674 (2009).

\bibitem{DZhiggs}
\DZero Collaboration, ``Combined upper limits on standard model Higgs 
boson production from the D\O\ experiment with up to 4.2 fb$^{-1}$ of data'' D\O\ Conference Note 5896 (2009).

\bibitem{CDFHiggsICHEP} CDF Collaboration, ``Combined Upper Limit on
Standard Model Higgs Boson Production for Summer 2008'', CDF Conference Note 9502 (2008).

\bibitem{DZHiggsICHEP} 
\DZero Collaboration, ``Combined upper limits on standard model Higgs 
boson production from the D\O\ experiment in 1.1-3.0 fb$^{-1}$'' D\O\ Conference Note 5756 (2008).

\bibitem{TevHiggsICHEP} The CDF and D\O\ Collaborations and the TEVNPHWG Working Group,
``Combined CDF and D\O\ Upper Limits on Standard Model Higgs Boson Production at Higg
Mass (155-200 GeV) with 3~fb$^{-1}$ of Data'', 
FERMILAB-PUB-08-270-E, CDF Note 9465, D\O\ Note 5754,
arXiv:0808.0534v1 [hep-ex] (2008).

\bibitem{tev-apr-2008} The CDF and D\O\ Collaborations and the TEVNPHWG Working Group,
``Combined CDF and D0 Upper Limits on Standard Model Higgs Boson Production with up to 2.4 fb$^{-1}$ of Data'',
FERMILAB-PUB-08-069-E, CDF Note 9290, D\O\ Note 5645,
arXiv:0804.3423v1 [hep-ex] (2008).


\bibitem{cdfWH} CDF Collaboration, ``Search for a Higgs Boson Produced in Association with a  \boldmath{$W$}
 Boson in \boldmath{$p\bar{p}$} Collisions at \boldmath{$\sqrt{s} = 1.96$} TeV'', arXiv:0906.5613,
Phys. Rev. Lett 103, 092002 (2009).


\bibitem{cdfZH}  CDF Collaboration, 
``Search for the Standard Model Higgs Boson in the \MET\ Plus Jets Sample'',
CDF Conference Note 9642 (2008).

\bibitem{cdfZHll} CDF Collaboration, ``A Search for $ZH\rightarrow\ell^+\ell^-b{\bar{b}}$ in 2.7~fb$^{-1}$
using a Neural Network Discriminant'', CDF Conference Note 9665 (2009).

\bibitem{cdfHWW} CDF Collaboration, ``Search for $H \rightarrow WW^*$ Production Using 3.6~fb$^{-1}$ of Data'',
CDF Conference Note 9500 (2009).

\bibitem{cdfHtt} CDF Collaboration, ``Search for SM Higgs using tau leptons using 2~fb$^{-1}$'',
CDF Conference Note 9179.

\bibitem{cdfjjbb} CDF Collaboration, ``A Search for the Standard Model Higgs Boson in the
All-Hadronic channel using a Matrix Element Method'',  CDF Conference Note 9366.

\bibitem{dzWHl} D\O\ Collaboration, ``Search for WH associated production using a combined Neural Network and
Matrix Element approach with 2.7 fb$^{-1}$ of Run II data,'' D\O\ Conference Note 5828.

\bibitem{dzWHt} D\O\ Collaboration, ``Search for the standard model Higgs boson in 
$\tau$ final states'', D\O\  Conference note 5883.

\bibitem{dzZHv} D\O\ Collaboration, ``Search for the standard model Higgs boson in the $HZ \rightarrow b \bar{b} \nu 
\nu$ channel in 2.1 fb$^{-1}$ of $p\bar{p}$ collisions at 
$\sqrt{s}=1.96$~TeV'', D\O\  Conference note 5586.


\bibitem{dzZHll} D\O\ Collaboration, ``A Search for \ZHll\  Production at
 D\O\ in \pp\ Collisions at $\sqrt{s}=1.96$~TeV'',
D\O\ Conference Note 5570.



%



\bibitem{dzWWW} D\O\ Collaboration, ``Search for associated Higgs boson
production $WH\rightarrow WWW^* \rightarrow \ell^\pm \nu
\ell^{\prime\pm} \nu^\prime +X$ in $p\bar{p}$ collisions at
$\sqrt{s}=1.96$~TeV'', 
D\O\ Conference Note 5485.

\bibitem{dzHWW} D\O\ Collaboration, ``Search 
for Higgs production in dilepton plus missing energy final
states with 3.0--4.2 fb$^{-1}$ of $p\overline{p}$
collisions at $\sqrt{s} =1.96$ TeV'',
D\O\ Conference Note 5871. 






\bibitem{dzHgg} D\O\ Collaboration, ``Search for the Standard Model Higgs boson
in   $\gamma \gamma$ final state with 4.2 fb$^{-1}$ data'', 
D\O\ Conference Note 5858.

\bibitem{dzttH} D\O\ Collaboration, ``Search for the standard model Higgs boson in the $
    t \bar{t} H \rightarrow t \bar{t}  b \bar{b} $ channel'', 
D\O\  Conference note 5739.

\bibitem{NEAT} K.~O.~Stanley and R.~Miikkulainen, ``Evolutionary Computation'', {\bf 10 (2)} 99-127 (2002); \\
S. Whiteson and D. Whiteson, hep-ex/0607012 (2006). 

\bibitem{pythia} 
T.~Sj\"ostrand, L.~Lonnblad and S.~Mrenna,
   ``PYTHIA 6.2: Physics and manual,''
  arXiv:hep-ph/0108264.

\bibitem{cteq} 
H.~L.~Lai {\it et al.}, ``Improved Parton
Distributions from Global Analysis of Recent Deep Inelastic Scattering
and Inclusive Jet Data'', Phys. Rev D \textbf{55}, 1280 (1997).

\bibitem{anastasiou}
  C.~Anastasiou, R.~Boughezal and F.~Petriello,
  ``Mixed QCD-electroweak corrections to Higgs boson production in gluon fusion'',
  arXiv:0811.3458 [hep-ph] (2008).

\bibitem{grazzinideflorian} D.~de~Florian and M.~Grazzini, ``Higgs production through gluon fusion:  updated
cross sections at the Tevatron and the LHC'', arXiv:0901.2427v1 [hep-ph] (2009).

\bibitem{nnlo1} 
S.~Catani, D.~de Florian, M.~Grazzini and P.~Nason,
   ``Soft-gluon resummation for Higgs boson production at hadron colliders,''
  JHEP {\bf 0307}, 028 (2003)
  [arXiv:hep-ph/0306211].

\bibitem{nnlo2} 
K.~A.~Assamagan {\it et al.}  [Higgs Working Group Collaboration],
   ``The Higgs working group: Summary report 2003,''
  arXiv:hep-ph/0406152.

\bibitem{aglietti} U. Aglietti, R. Bonciani, G. Degrassi, A. Vicini, ``Two-loop electroweak corrections to Higgs 
production in proton-proton collisions'', arXiv:hep-ph/0610033v1 (2006).
\bibitem{mstw2008}   A.~D.~Martin, W.~J.~Stirling, R.~S.~Thorne and G.~Watt,
  ``Parton distributions for the LHC'',
  arXiv:0901.0002 [hep-ph] (2009).

\bibitem{mrst2002}
 A.~D.~Martin, R.~G.~Roberts, W.~J.~Stirling and R.~S.~Thorne,
  Phys.\ Lett.\  B {\bf 531}, 216 (2002)
  [arXiv:hep-ph/0201127].


\bibitem{Brein}
O. Brein, A. Djouadi, and R. Harlander,
``NNLO QCD corrections to the Higgs-strahlung processes at
                  hadron  colliders'',
Phys. Lett. B579, 2004, 149-156.

\bibitem{Ciccolini}
Ciccolini, M. L. and Dittmaier, S. and Kramer, M.,
``Electroweak radiative corrections to associated W H and Z
                  H production  at hadron colliders'',
Phys. Rev. D68 (2003) 073003.


\bibitem{Berger}
E. Berger and J. Campbell.
 ``Higgs boson production in weak boson fusion at next-to-leading order'',
 Phys. Rev. D70 (2004) 073011,


\bibitem{hdecay}
A.~Djouadi, J.~Kalinowski and M.~Spira,
   ``HDECAY: A program for Higgs boson decays in the standard model and its
   supersymmetric extension,''
  Comput.\ Phys.\ Commun.\  {\bf 108}, 56 (1998)
  [arXiv:hep-ph/9704448].

\bibitem{alpgen}
M.~L.~Mangano, M.~Moretti, F.~Piccinini, R.~Pittau and A.~D.~Polosa,
   ``ALPGEN, a generator for hard multiparton processes in hadronic
   collisions,''
  JHEP {\bf 0307}, 001 (2003)
  [arXiv:hep-ph/0206293].

\bibitem{MC@NLO}
S. Frixione and B.R. Webber, 
JHEP 06, 029 (2002)   [arXiv:hep-ph/0204244]

\bibitem{herwig} 
G.~Corcella {\it et al.},
   ``HERWIG 6: An event generator for hadron emission reactions with
   interfering gluons (including supersymmetric processes),''
  JHEP {\bf 0101}, 010 (2001)
  [arXiv:hep-ph/0011363].

\bibitem{comphep}
A.~Pukhov {\it et al.},
   ``CompHEP: A package for evaluation of Feynman diagrams and integration  over
   multi-particle phase space. User's manual for version 33,''
  [arXiv:hep-ph/9908288].

\bibitem{mcfm} J.~Campbell and R.~K.~Ellis, 
 http://mcfm.fnal.gov/. 
%


\bibitem{pdgstats}
T. Junk, Nucl. Instrum. Meth. A434, p. 435-443, 1999,
A.L.~Read, ``Modified frequentist analysis of search results (the $CL_s$ method)'', in                              
F.~James, L.~Lyons and Y.~Perrin (eds.), {\sl Workshop on Confidence Limits},                                   
CERN, Yellow Report 2000-005, available through {\tt cdsweb.cern.ch}.

\bibitem{pflh} W.~Fisher, ``Systematics and Limit Calculations,''
FERMILAB-TM-2386-E.



















\end{thebibliography}

\end{document}